\documentclass[11pt]{article}

\usepackage[vmargin=1in]{geometry}
\usepackage[T1]{fontenc}
\usepackage{amsmath,amssymb}
\usepackage{hyperref}

\newcommand{\dd}{\mathrm{d}}
\newcommand{\X}{\mathfrak{X}}
\newcommand{\bx}{\mathbf{x}}
\newcommand{\bu}{\mathbf{u}}

\newcommand{\bxi}{\boldsymbol{\xi}}
\renewcommand{\iota}{\mathbf{i}}

\title{Relabeling symmetry and conservation laws in hydrostatic
and nonhydrostatic upper-ocean models}

\author{F.J.\ Beron-Vera\\
Department of Atmospheric Sciences\\
Rosenstiel School of Marine, Atmospheric \& Earth Science\\
University of Miami\\
Miami, FL 33145\\
\href{mailto:fberon@miami.edu}{fberon@miami.edu}
}

\date{Started: April 25, 2023. This version: \today.}

\begin{document}

\maketitle

\begin{abstract}
     We explicitly identify conserved integrals of hydrostatic and nonhydrostatic reduced-gravity upper-ocean models, both vertically unrestricted and columnar, with Noether quantities associated with particle relabeling. The columnar velocity reduction preserves the Lie-algebra structure, providing a common Euler--Poincar\'e framework for both families. Within this framework, we introduce nonhydrostatic columnar models with polynomial vertical buoyancy structure, extending the hydrostatic stratified hierarchy. Retaining the temporal endpoint pairing, we construct transported relabeling generators and evaluate their vorticity-dependent conserved integrals, assuming that spatial boundary terms vanish. For stratified columnar models, preservation of the effective buoyancy entering the Lagrangian suffices for symmetry even when individual buoyancy coefficients change, yielding conserved quantities that generally are not Casimirs of the full Lie--Poisson bracket. Where the Legendre transform is regular, we state the corresponding brackets and identify their Casimir families. The hydrostatic parent is treated directly through the Euler--Poincar\'e principle because its Legendre transform is degenerate. Integrals following directly from advection are distinguished from those obtained through the Noether pairing.
\end{abstract}

\tableofcontents

\section{Introduction}
\label{sec:introduction}

Particle relabeling symmetry provides a variational explanation for fundamental circulation and vorticity conservation laws in ideal fluids. Its connection with fluid invariants has a substantial history, including the early variational treatments of \cite{Eckart-60,Eckart-63}, the exchange-invariance formulation of \cite{Newcomb-67}, and the geophysical and continuum-mechanical developments of \cite{Ripa-AIP-81,Henyey-82,Salmon-82,Padhye-Morrison-96a}. Systematic accounts are given by \cite{Salmon-88a,Salmon-98}. These developments show how appropriately specified transformations of particle labels generate conservation laws through Noether's theorem \cite{Noether-18}.

The conserved quantity appears as the temporal boundary term in the first variation of the action. For a reduced fluid Lagrangian $\ell(u,a)$, where $u$ is the Eulerian velocity and $a$ denotes the advected quantities, the Euler--Poincar\'e constrained variations yield the endpoint pairing
\begin{equation}
  J_\xi:=\left\langle\frac{\delta\ell}{\delta u},\xi\right\rangle,
\end{equation}
where $\xi$ is the infinitesimal displacement generating the variation. A variational symmetry makes this pairing constant along solutions. We retain the temporal endpoint term while assuming that spatial boundary contributions vanish under suitable conditions on the fields and admissible variations.

An explicit application of this argument to fluid vorticity is provided by \cite{Samelson-Samelson-05}, who construct volume-preserving, entropy-preserving parcel exchanges and obtain Ertel potential-vorticity conservation from the Noether pairing. The Euler--Poincar\'e framework of \cite{Holm-etal-98a,Holm-etal-02} supplies a general Eulerian formulation for fluids with advected quantities. Within this framework, \cite{Cotter-Holm-12} develop the transport equations for relabeling generators and the conditions under which they preserve the advected fields. More recently, \cite{Beron-Luesink-25} explicitly identify conserved functionals of stratified thermal quasigeostrophic flow with Noether quantities. Here we construct generators and evaluate their pairings for vertically unrestricted upper-ocean models and their columnar reductions, in both hydrostatic and nonhydrostatic settings.

The physical configuration consists of an active fluid layer bounded above by a horizontal rigid lid and below by a material interface with an infinitely deep, quiescent fluid. This reduced-gravity setting provides a useful first approximation to upper-ocean dynamics, reflecting the concentration of variability toward the surface \cite{Gill-82}. Homogeneous-layer shallow-water dynamics can be extended within this setting to include horizontally varying buoyancy and vertical buoyancy structure \cite{Ripa-GAFD-93,Ripa-JFM-95,Beron-21-POFb}. We consider the hydrostatic primitive-equation (PE) and nonhydrostatic Euler--Boussinesq (EB) branches. Their vertically unrestricted parents are denoted IL$^\infty$PE and IL$^\infty$EB. Their columnar reductions are denoted HLPE and HLEB for homogeneous buoyancy, IL$^0$PE and IL$^0$EB for vertically uniform but horizontally varying buoyancy, and IL$^{(0,\alpha)}$PE and IL$^{(0,\alpha)}$EB for polynomial vertical buoyancy structure. The nonhydrostatic columnar models belong to the Green--Naghdi class \cite{Green-Naghdi-76,Percival-etal-08,Holm-Street-26}. The stratified IL$^{(0,\alpha)}$EB family introduced here extends the hydrostatic hierarchy of \cite{Beron-21-POFb} by retaining vertical kinetic energy.

A central distinction concerns what a relabeling transformation must preserve. Preservation of velocity and every advected field is sufficient for invariance of the reduced Lagrangian, but it is not necessary. In the stratified columnar models, the buoyancy coefficients are individually advected, whereas the Lagrangian depends on them only through the effective buoyancy. A transported displacement can preserve the thickness measure and effective buoyancy without preserving each coefficient separately. It therefore generates a conserved Noether quantity even when some advected fields change under the variation. The resulting vorticity-dependent integrals are \emph{generally not Casimirs of the full Lie--Poisson bracket}: their conservation depends on the particular Lagrangian.

We make these statements explicit by verifying that the columnar velocity reduction preserves the Lie-algebra structure, constructing transported relabeling generators, checking Lagrangian preservation, and evaluating their endpoint pairings. The hydrostatic and nonhydrostatic reductions admit a common treatment, with their different kinetic energies entering through the corresponding momenta and potential vorticities. For the unrestricted parents, the construction uses the momentum one-form and Ertel potential vorticity. Where the Legendre transform is regular, we state the corresponding Lie--Poisson brackets and identify their Casimir families alongside the Noether constructions. The hydrostatic parent is treated directly through the Euler--Poincar\'e principle because its full velocity Legendre transform is degenerate. Throughout, we distinguish conservation by relabeling symmetry, conservation by material advection, and Casimir status. The implications of retaining spatial boundary contributions are discussed at the end.

\section{Euler--Poincar\'e principle and relabeling symmetry}
\label{sec:EP}

\subsection{Advected quantities and constrained variations}

Let $u=u(x,t)$ denote a time-dependent vector field on the fluid domain $M$, with $u(\cdot,t)\in\X(M)$. In Cartesian coordinates we write $u=\bu(x,t)\cdot\nabla$. The discussion applies in $n=2$ or $3$ spatial dimensions. Spatial and temporal dependence will generally be suppressed below. As usual in the Euler--Poincar\'e formulation \cite{Holm-etal-02}, we assume sufficient regularity and appropriate boundary or asymptotic conditions for all functionals and integrations by parts to be well defined. We assume that all spatial boundary terms arising in first variations, duality identities, and evaluations of the Noether pairing vanish. An admissible functional has well-defined spatial integral derivatives belonging to the chosen velocity and advected-field spaces, with the same convention for integrations by parts. An admissible displacement belongs to the velocity Lie algebra. These assumptions apply also to the lid and material interface in the parent models; horizontal decay or periodicity alone does not enforce them there. We retain temporal endpoint contributions, which are the source of the Noether quantities. A characterization of boundary phase spaces is beyond the scope of this work and is discussed in Section~\ref{sec:discussion}.

Let $V(M)$ denote a representation space of the diffeomorphism group of $M$, and let $V(M)^*$ denote its dual with respect to the appropriate spatial integral pairing. An advected quantity $a$, with $a(\cdot,t)\in V(M)^*$, satisfies
\begin{equation}
  (\partial_t+\pounds_u)a=0.
  \label{eq:advection}
\end{equation}
Here $\pounds_u$ is the Lie derivative, whose form depends on the tensor character of the advected quantity. For an advected scalar $s$ and an advected density $\rho\,\dd^nx$, respectively,
\begin{equation}
  \partial_t s+\bu\cdot\nabla s=0,
  \qquad
  \partial_t\rho+\nabla\cdot\rho\bu=0.
  \label{eq:scalar-density-advection}
\end{equation}
The symbol $\dd^nx$ denotes the Cartesian area or volume measure, identified with the standard oriented form $\dd x^1\wedge\cdots\wedge\dd x^n$ when using differential forms. We allow a finite collection $a=(a_1,\ldots,a_N)$, where $N\ge1$ is an integer and $a_i(\cdot,t)\in V_i(M)^*$, with $V(M)^*=\bigoplus_{i=1}^N V_i(M)^*$. Each member is transported according to its tensor character.

The constrained variations in the Euler--Poincar\'e principle are
\begin{equation}
  \delta u=\partial_t\xi+[u,\xi],
  \qquad
  \delta a_i=-\pounds_\xi a_i,
  \qquad i=1,\ldots,N.
  \label{eq:EP-variations}
\end{equation}
The displacement $\xi=\bxi\cdot\nabla$ is the Eulerian representation of an infinitesimal variation of fluid trajectories. The Lie bracket on $\X(M)$ is given by the commutator of vector fields, which in Cartesian coordinates reads
\begin{equation}
  [u,\xi]
  =
  \left(
    (\bu\cdot\nabla)\bxi
    -
    (\bxi\cdot\nabla)\bu
  \right)\cdot\nabla.
  \label{eq:vector-bracket}
\end{equation}
Thus, for the scalar and density examples,
\begin{equation}
  \delta s=-\bxi\cdot\nabla s,
  \qquad
  \delta\rho=-\nabla\cdot\rho\bxi.
  \label{eq:scalar-density-variations}
\end{equation}
These variations express the displacement of advected fields with their reference values held fixed.

Let the reduced Lagrangian be $\ell(u,a_1,\ldots,a_N)$ and define
\begin{equation}
  m:=\frac{\delta\ell}{\delta u},
  \qquad
  b_i:=\frac{\delta\ell}{\delta a_i},
  \qquad i=1,\ldots,N.
  \label{eq:momentum-and-derivatives}
\end{equation}
Thus $m\in\X(M)^*$ and $b_i\in V_i(M)$, where the duals are understood with respect to the spatial integral pairings introduced below. The functional derivatives are defined through
\begin{equation}
  \delta\ell
  =
  \langle m,\delta u\rangle
  +
  \sum_{i=1}^{N}\langle b_i,\delta a_i\rangle.
  \label{eq:first-variation-Lagrangian}
\end{equation}
The notation $\langle\, ,\hspace{.01cm}\rangle$ denotes the appropriate spatial integral pairing. We use the $L^2$ pairing to represent functional derivatives, formally identifying the corresponding Hilbert-space duals through the Riesz representation theorem under suitable regularity assumptions. In particular, momentum is a one-form density,
\begin{equation}
  m=\mathbf{m}\cdot\dd x\otimes\dd^nx,
  \qquad
  \langle m,\xi\rangle
  =
  \int_M\mathbf{m}\cdot\bxi\,\dd^nx.
\end{equation}
The boldface $\mathbf{m}$ denotes the coordinate coefficients of $m$, rather than an identification of momentum with velocity.

Integration by parts gives the duality identity
\begin{equation}
  \langle m,[u,\xi]\rangle
  =
  -\langle\pounds_u m,\xi\rangle,
  \label{eq:bracket-duality}
\end{equation}
where the coordinate representation of the Lie derivative of a one-form density is
\begin{equation}
  \pounds_{\bu}\mathbf{m}
  =
  (\bu\cdot\nabla)\mathbf{m}
  +
  (\nabla\bu)\mathbf{m}
  +
  \mathbf{m}\,\nabla\cdot\bu.
  \label{eq:momentum-Lie-derivative}
\end{equation}
Here the velocity gradient is defined by $(\nabla\bu)_{jk}=\partial_j u^k$, so $(\nabla\bu)\mathbf{m} = \sum_j m_j\nabla u^j$.

We use the diamond notation to collect the contributions from the advected quantities:
\begin{equation}
  \langle b_i\diamond a_i,\xi\rangle
  :=
  -\langle b_i,\pounds_\xi a_i\rangle.
  \label{eq:diamond-definition}
\end{equation}
For example, if $a_i=\rho\,\dd^nx$ and $b_i=s$ is a scalar, then
\begin{equation}
  b_i\diamond a_i=\rho\,\dd s\otimes\dd^nx.
\end{equation}
If $a_i=\tilde s$ is a scalar and $b_i=\tilde\rho\,\dd^nx$, then
\begin{equation}
  b_i\diamond a_i=-\tilde\rho\,\dd\tilde s\otimes\dd^nx.
\end{equation}
Their Cartesian coefficients are $\rho\nabla s$ and $-\tilde\rho\nabla\tilde s$, respectively.

\subsection{The endpoint pairing}

Consider the action
\begin{equation}
  \mathfrak S
  =
  \int_{t_0}^{t_1}\ell(u,a_1,\ldots,a_N)\,\dd t.
\end{equation}
Substituting the constrained variations gives
\begin{equation}
  \delta\mathfrak S
  =
  \int_{t_0}^{t_1}
  \left(
    \langle m,\partial_t\xi+[u,\xi]\rangle
    -
    \sum_{i=1}^{N}\langle b_i,\pounds_\xi a_i\rangle
  \right)\dd t.
\end{equation}
Integrating the time derivative by parts, using \eqref{eq:bracket-duality}, and applying the definition of the diamond operator, we obtain
\begin{equation}
  \delta\mathfrak S
  =
  \left.\langle m,\xi\rangle\right|_{t_0}^{t_1}
  -
  \int_{t_0}^{t_1}
  \left\langle
    (\partial_t+\pounds_u)m
    -
    \sum_{i=1}^{N}b_i\diamond a_i,
    \xi
  \right\rangle\,\dd t.
  \label{eq:action-endpoint-identity}
\end{equation}
Variations supported in the temporal interior yield the Euler--Poincar\'e equation
\begin{equation}
  (\partial_t+\pounds_u)m
  =
  \sum_{i=1}^{N}b_i\diamond a_i,
  \label{eq:general-EP}
\end{equation}
together with the advection equations \eqref{eq:advection} for $a_1,\ldots,a_N$.

We retain the endpoint pairing and define
\begin{equation}
  J_\xi:=\langle m,\xi\rangle.
  \label{eq:Noether-pairing}
\end{equation}
Retaining this term does not mean imposing stationarity under arbitrary independent endpoint displacements. Rather, it allows the consequences of a symmetry to be read from the first-variation identity.

Indeed, on an Euler--Poincar\'e solution,
\begin{equation}
  \frac{\dd J_\xi}{\dd t}
   =
  \langle\partial_t m,\xi\rangle
  +
  \langle m,\partial_t\xi\rangle
  =
  \langle m,\partial_t\xi+[u,\xi]\rangle
  -
  \sum_{i=1}^{N}\langle b_i,\pounds_\xi a_i\rangle
  =
  \delta_\xi\ell.
  \label{eq:Noether-identity}
\end{equation}
Here $\delta_\xi\ell$ denotes the variation induced by \eqref{eq:EP-variations}. Thus a constrained variation for which $\delta_\xi\ell=0$ gives a conserved $J_\xi$. We consider exact invariance here; invariance up to a total derivative requires the corresponding correction to the Noether quantity.

\subsection{Two symmetry conditions}

Particle relabeling acts on the flow map $\varphi_t$ by composition on the right, $\varphi_t\mapsto\varphi_t\circ r$, where $r$ is a time-independent diffeomorphism of the reference domain. Depending on how the reference fields are treated, this action describes either a change of particle names or an exchange of parcels carrying prescribed material properties. Writing $u=\dot\varphi_t\circ\varphi_t^{-1}$, the Eulerian velocity is unchanged:
\begin{equation}
  u
  \mapsto
  \dot\varphi_t\circ r\circ(\varphi_t\circ r)^{-1}
  =
  \dot\varphi_t\circ\varphi_t^{-1}
  =
  u.
\end{equation}

Let $a_0$ denote the prescribed reference fields, with Eulerian distributions $a_t=(\varphi_t)_*a_0$, where pushforward is understood according to the tensor character of each field. We consider exchanges with $a_0$ held fixed: only the flow map is transformed, while the material properties assigned to each reference parcel remain unchanged. Consequently,
\begin{equation}
  a_t
  \mapsto
  (\varphi_t\circ r)_*a_0
  =
  (\varphi_t)_*(r_*a_0).
\end{equation}
The Eulerian advected fields are therefore unchanged if and only if $r_*a_0=a_0$. Beyond time independence, \emph{this generally imposes a further restriction on $r$}. For a scalar, only exchanges between reference locations with the same value are allowed; a uniform reference scalar imposes no such restriction. For a density, even a uniform reference distribution requires volume-preserving transformations.

Let $\xi$ be the Eulerian displacement induced by a one-parameter family of these time-independent right compositions. With $a_0$ held fixed, the induced variations satisfy
\begin{equation}
  \delta_\xi u
  =
  \partial_t\xi+[u,\xi]
  =
  0,
  \qquad
  \delta_\xi a_i
  =
  -\pounds_\xi a_i.
  \label{eq:transported-generator}
\end{equation}
Thus transport of the generator ensures preservation of the velocity, while preservation of each advected field requires the additional condition $\pounds_\xi a_i=0$. Substitution into \eqref{eq:Noether-identity} gives
\begin{equation}
  \frac{\dd J_\xi}{\dd t}
  =
  -\sum_{i=1}^{N}
  \left\langle
    \frac{\delta\ell}{\delta a_i},
    \pounds_\xi a_i
  \right\rangle.
  \label{eq:transported-Noether-identity}
\end{equation}

We distinguish two conditions under which the right-hand side vanishes. The stronger condition is preservation of every advected quantity:
\begin{equation}
  \pounds_\xi a_i=0,
  \qquad i=1,\ldots,N.
  \label{eq:individual-preservation}
\end{equation}
These generators correspond to the subgroup of right actions preserving all the reference fields, $r_*a_{i0}=a_{i0}$. For an advected density $\rho\,\dd^nx$ and an advected scalar $s$, the infinitesimal conditions are
\begin{equation}
  \nabla\cdot \rho\bxi =0,
  \qquad
  \bxi\cdot\nabla s=0.
\end{equation}
Together with $\delta_\xi u=0$, they ensure invariance of any reduced Lagrangian of the stated variables.

The weaker condition requires preservation only of the particular Lagrangian:
\begin{equation}
  \delta_\xi\ell
  =
  -\sum_{i=1}^{N}
  \left\langle
    \frac{\delta\ell}{\delta a_i},
    \pounds_\xi a_i
  \right\rangle
  =
  0.
  \label{eq:Lagrangian-preservation}
\end{equation}
Individual advected fields may then change, provided their combined contribution to the Lagrangian variation vanishes.

For example, take $a_1=\rho\,\dd^nx$ and $a_k=s_k$, $k=2,\ldots,N$, and suppose that the Lagrangian depends on the advected scalars only through
\begin{equation}
  \ell=\ell(u,\rho,S),
  \qquad
  S=\sum_{k=2}^{N}c_k s_k,
\end{equation}
with constant coefficients $c_k$. A transported displacement satisfying
\begin{equation}
  \delta_\xi\rho=0,
  \qquad
  \delta_\xi S
  =
  \sum_{k=2}^{N}c_k\delta_\xi s_k
  =
  0
\end{equation}
preserves the Lagrangian even when some $\delta_\xi s_k\ne0$. It preserves the combination entering the dynamics without necessarily preserving each constituent field.

The first condition gives symmetries shared by arbitrary Lagrangians of the specified advected variables. The second allows additional symmetries associated with the particular dependence of the Lagrangian on those variables. The resulting Noether quantities need not be Casimirs of the full Lie--Poisson bracket, when such a formulation is available. In either case, an admissible transported generator and the relevant invariance condition must be established before evaluating the Noether quantity \eqref{eq:Noether-pairing}. Integrals following directly from advection will be identified separately.

\subsection{Lie--Poisson duality and Casimirs}
\label{sec:LP-Casimirs}

For models admitting a Hamiltonian formulation dual to the Euler--Poincar\'e formulation, the partial Legendre transform with respect to $u$ defines the Hamiltonian
\begin{equation}
  \mathcal H(m,a)=\langle m,u\rangle-\ell(u,a),
  \label{eq:Legendre-transform}
\end{equation}
where $u$ is understood to be expressed in terms of $m$ and $a$.

The Hamiltonian variables $\mu=(m,a)$ belong to the dual of the semidirect-product Lie algebra $\mathfrak s=\X(M)\ltimes V(M)$, namely, $\mathfrak s^*\simeq\X(M)^*\times V(M)^*$. This structure reflects the action of the diffeomorphism group on the advected quantities and leads to the Lie--Poisson formulation dual to the Euler--Poincar\'e equations \cite{Holm-etal-02}.

The evolution of an arbitrary admissible functional $\mathcal F(\mu)$ is given by
\begin{equation}
  \frac{\dd\mathcal F}{\dd t}
  =\{\mathcal F,\mathcal H\},
  \label{eq:LP-evolution}
\end{equation}
where the Lie--Poisson bracket is
\begin{equation}
  \{\mathcal F,\mathcal G\}
  =
  -\left\langle m,\left[\frac{\delta\mathcal F}{\delta m},\frac{\delta\mathcal G}{\delta m}\right]\right\rangle
  +
  \sum_{i=1}^{N}\left\langle a_i,\pounds_{\frac{\delta\mathcal G}{\delta m}}\frac{\delta\mathcal F}{\delta a_i}-\pounds_{\frac{\delta\mathcal F}{\delta m}}\frac{\delta\mathcal G}{\delta a_i}\right\rangle.
  \label{eq:general-LP-bracket}
\end{equation}
The Lie derivatives in the second term act on the representation spaces dual to the advected quantities. With the commutator convention \eqref{eq:vector-bracket}, the displayed signs give $(\partial_t+\pounds_u)m=-\sum_i(\delta\mathcal H/\delta a_i)\diamond a_i$ and $u=\delta\mathcal H/\delta m$, consistently with \eqref{eq:general-EP} and $\delta\mathcal H/\delta a_i=-\delta\ell/\delta a_i$.

A Casimir $\mathcal C(\mu)$ belongs to the kernel of the Lie--Poisson bracket, namely,
\begin{equation}
  \{\mathcal F,\mathcal C\}=0
  \,\text{for every admissible }\mathcal F.
  \label{eq:Casimir-definition}
\end{equation}
Consequently, Casimirs are conserved independently of the particular Hamiltonian.

Our objective is to identify these Casimirs with the Noether quantities $J_\xi$ arising from particle-relabeling symmetry. In particular, we seek explicit generators $\xi$ satisfying \eqref{eq:transported-generator} and \eqref{eq:individual-preservation} such that
\begin{equation}
  J_\xi=\langle m,\xi\rangle
  \longleftrightarrow\mathcal C,
  \label{eq:Noether-Casimir-correspondence}
\end{equation}
where the correspondence may involve separately conserved contributions from advected fields. The identification is made for each model within the common class of admissible fields and functionals for which spatial boundary terms vanish.

The stronger relabeling condition \eqref{eq:individual-preservation} is independent of the particular Lagrangian and is therefore the natural starting point for identifying Casimirs. By contrast, the weaker condition \eqref{eq:Lagrangian-preservation} may yield Noether quantities whose conservation depends on the chosen Hamiltonian and which need not be Casimirs of the full Lie--Poisson bracket.

The Euler--Poincar\'e construction does not require an invertible Legendre transform. This distinction is relevant for IL$^\infty$PE: its Lagrangian has no vertical kinetic energy, so its momentum does not determine the full velocity. We work directly with its Euler--Poincar\'e equations and particle-relabeling symmetries without inferring a regular Lie--Poisson dual by Legendre transformation.

\section{Reduced-gravity models and their momenta}
\label{sec:models}

We consider an active layer of inviscid fluid bounded above by a horizontal rigid lid at $z=0$ and below by a material interface at $z=-h(\bx,t)$, where $h>0$ is the layer thickness. The fluid beneath the interface is homogeneous, infinitely deep, and quiescent, with density $\rho_{\mathrm{down}}$. This is the reduced-gravity, or one-and-a-half-layer, setting. Let $\mathcal D$ denote the horizontal domain. The active fluid occupies
\begin{equation}
  M_t
  =
  \left\{
    x=(\bx,z):
    \bx\in\mathcal D,\;
    -h(\bx,t)<z<0
  \right\}.
\end{equation}
We use Cartesian coordinates $x=(\bx,z)$, with $\bx=(x^1,x^2)$ and $z$ increasing upward. The standard orientations are $\dd^2\bx=\dd x^1\wedge\dd x^2$ and $\dd^3x=\dd^2\bx\wedge\dd z=\dd x^1\wedge\dd x^2\wedge\dd z$. For integration, we suppress the wedge and write $\dd^3x=\dd^2\bx\,\dd z$. Below, $\nabla$ denotes the horizontal gradient and $\nabla_3=(\nabla,\partial_z)$ the full spatial gradient.

Let $\bu(\bx,z,t)$ and $w(\bx,z,t)$ denote the horizontal and vertical velocities, respectively. The full velocity is $\bu_3=(\bu,w)$, and the material derivative is
\begin{equation}
  D_t
  :=
  \partial_t+\bu\cdot\nabla+w\partial_z.
\end{equation}
In the Boussinesq approximation, the buoyancy relative to the lower fluid is
\begin{equation}
  \vartheta(\bx,z,t)
  :=
  -g\frac{\rho_{\mathrm{up}}(\bx,z,t)-\rho_{\mathrm{down}}}{\rho_{\mathrm{ref}}},
\end{equation}
where $\rho_{\mathrm{up}}$ is the active-layer density, $g>0$ is gravitational acceleration, and $\rho_{\mathrm{ref}}$ is a constant reference density. Thus $\vartheta>0$ corresponds to active fluid lighter than the lower fluid. All Lagrangians and energies below are divided by $\rho_{\mathrm{ref}}$.

The active fluid is incompressible and buoyancy is materially conserved:
\begin{equation}
  \nabla\cdot\bu+\partial_z w=0,
  \qquad
  D_t\vartheta=0.
  \label{eq:parent-kinematics}
\end{equation}
The rigid lid and material interface impose
\begin{equation}
  w=0
  \quad\hbox{at }z=0,
  \qquad
  w=-\partial_t h-\bu\cdot\nabla h
  \quad\hbox{at }z=-h.
  \label{eq:vertical-boundary-conditions}
\end{equation}
At solid lateral boundaries, the horizontal velocity has vanishing normal component. Periodic domains or suitable decay conditions may be considered instead.

Let $P$ denote pressure measured relative to a fixed atmospheric pressure, and define the pressure relative to the hydrostatic lower fluid by
\begin{equation}
  p
  :=
  \frac{P+g\rho_{\mathrm{down}}z}{\rho_{\mathrm{ref}}}.
\end{equation}
Pressure continuity across the interface with the quiescent lower fluid gives
\begin{equation}
  p=0
  \quad\hbox{at }z=-h,
  \label{eq:interface-pressure}
\end{equation}
after choosing the additive pressure constant. There is no mass exchange across this interface.

\subsection{The primitive-equation and Euler--Boussinesq parents}

We begin with the unrestricted models IL$^\infty$PE and IL$^\infty$EB. Let $\mathbf f(\bx)$ be a horizontal vector potential for the vertical component of planetary rotation, satisfying
\begin{equation}
  \hat{\mathbf z}\cdot\nabla\times\mathbf f=f,
  \label{eq:rotation-potential}
\end{equation}
where $f=f(x^2)$ is the Coriolis parameter. We adopt the traditional approximation, retaining only the vertical component of planetary rotation.

To formulate both parents simultaneously, introduce $\varepsilon\in\{0,1\}$, with $\varepsilon=0$ for PE and $\varepsilon=1$ for EB. Let $D$ denote the Jacobian of the inverse flow, regarded as the coefficient of an advected volume density $D\,\dd^3x$. Incompressibility is imposed by the constraint $D=1$. The reduced Lagrangian is
\begin{equation}
  \ell_\varepsilon(\bu_3,\vartheta,D;p)
  =
  \int_{M_t}
    D\left(
      \frac12|\bu|^2
      +\frac{\varepsilon}{2}w^2
      +\bu\cdot\mathbf f
      +\vartheta z
    \right)
    -p(D-1)
    \,\dd^3x.
  \label{eq:parent-Lagrangian}
\end{equation}
The pressure $p$ acts as a Lagrange multiplier enforcing incompressibility. The PE parent retains horizontal kinetic energy, whereas the EB parent includes the vertical kinetic energy.

The advected quantities are the volume density $D\,\dd^3x$ and the buoyancy scalar $\vartheta$. Their Euler--Poincar\'e constrained variations, together with that of the velocity, are
\begin{equation}
  \delta\bu_3
  =
  \partial_t\boldsymbol\xi_3
  +[\bu_3,\boldsymbol\xi_3]_3,
  \qquad
  \delta D=-\nabla_3\cdot D\boldsymbol\xi_3,
  \qquad
  \delta\vartheta=-\boldsymbol\xi_3\cdot\nabla_3\vartheta,
  \label{eq:parent-variations}
\end{equation}
where $[\,,\hspace{.01cm}]_3$ denotes the three-dimensional vector-field commutator, defined as in \eqref{eq:vector-bracket}. The displacement $\boldsymbol\xi_3$ respects the rigid-lid, material-interface, and lateral boundary conditions.

The functional derivatives of \eqref{eq:parent-Lagrangian} are
\begin{equation}
  \frac{\delta\ell_\varepsilon}{\delta\bu_3}
  =
  D(\bu+\mathbf f,\varepsilon w),
  \qquad
  \frac{\delta\ell_\varepsilon}{\delta\vartheta}
  =Dz,
  \qquad
  \frac{\delta\ell_\varepsilon}{\delta D}
  =
  \frac12|\bu|^2+\frac{\varepsilon}{2}w^2
  +\bu\cdot\mathbf f+\vartheta z-p.
  \label{eq:parent-functional-derivatives}
\end{equation}
Substitution into the Euler--Poincar\'e equation \eqref{eq:general-EP}, followed by setting $D=1$, gives
\begin{subequations}
\label{eq:parent-equations}
\begin{align}
  D_t\bu+f\hat{\mathbf z}\times\bu
  &=-\nabla p,\\
  \varepsilon D_t w
  &=-\partial_z p+\vartheta,\\
  D_t\vartheta&=0,\\
  \nabla\cdot\bu+\partial_z w&=0.
\end{align}
\end{subequations}
For $\varepsilon=0$, the vertical momentum equation reduces to hydrostatic balance, $\partial_zp=\vartheta$. For $\varepsilon=1$, it retains vertical acceleration and gives the Euler--Boussinesq equations. Both systems satisfy the kinematic and pressure boundary conditions stated above.

\subsection{Columnar reduction and advected buoyancy moments}

We now restrict the horizontal velocity to be independent of depth,
\begin{equation}
  \bu(\bx,z,t)=\bar{\bu}(\bx,t),
  \label{eq:columnar-horizontal-velocity}
\end{equation}
where the overbar denotes the vertical average over $-h<z<0$. Incompressibility and the rigid-lid condition in \eqref{eq:parent-kinematics}--\eqref{eq:vertical-boundary-conditions} then give
\begin{equation}
  w=-z\nabla\cdot\bar{\bu}.
  \label{eq:columnar-vertical-velocity}
\end{equation}
Evaluating this expression at the material interface yields the thickness equation
\begin{equation}
  \partial_t h+\nabla\cdot h\bar{\bu}=0,
  \label{eq:reduced-continuity}
\end{equation}
or, equivalently, $\bar D_t h=-h\nabla\cdot\bar{\bu}$, where $\bar D_t:=\partial_t+\bar{\bu}\cdot\nabla$.

The columnar velocity is thus the three-dimensional divergence-free vector field
\begin{equation}
  u_3=\bar{\bu}\cdot\nabla-z(\nabla\cdot\bar{\bu})\partial_z.
  \label{eq:columnar-vector-field}
\end{equation}
For an arbitrary horizontal displacement $\bar\xi=\bar{\bxi}\cdot\nabla\in\X(\mathcal D)$, define its divergence-free lift by
\begin{equation}
  \xi_3=\bar{\bxi}\cdot\nabla-z(\nabla\cdot\bar{\bxi})\partial_z.
  \label{eq:columnar-displacement}
\end{equation}
Both vector fields satisfy the rigid-lid condition at $z=0$. Direct calculation gives
\begin{equation}
  [u_3,\xi_3]_3
  =
  [\bar u,\bar\xi]
  -z\nabla\cdot\left((\bar{\bu}\cdot\nabla)\bar{\bxi}-(\bar{\bxi}\cdot\nabla)\bar{\bu}\right)\partial_z,
  \label{eq:columnar-bracket-closure}
\end{equation}
where $[\bar u,\bar\xi]$ is the horizontal vector-field commutator. Thus the columnar vector fields are closed under the three-dimensional commutator and form a Lie subalgebra of the divergence-free vector fields. In other words, \emph{the columnar velocity reduction preserves the Lie-algebra structure}. Consequently, the horizontal component of the constrained variation $\delta u_3=\partial_t\xi_3+[u_3,\xi_3]_3$ reduces to
\begin{equation}
  \delta\bar u=\partial_t\bar\xi+[\bar u,\bar\xi].
  \label{eq:reduced-velocity-variation}
\end{equation}

To describe vertical buoyancy structure, we introduce the dimensionless coordinate following \cite{Ripa-JFM-95}:
\begin{equation}
  \sigma:=1+\frac{2z}{h},
  \qquad -1\le\sigma\le1.
  \label{eq:sigma-coordinate}
\end{equation}
Using \eqref{eq:columnar-vertical-velocity} and \eqref{eq:reduced-continuity}, one finds
\begin{equation}
  D_t\sigma
  =
  \frac{2w}{h}
  -\frac{2z}{h^2}\bar D_t h
  =0.
  \label{eq:sigma-advection}
\end{equation}
Thus $\sigma$ is materially conserved by the columnar flow.

Following \cite{Beron-21-POFb}, where the hydrostatic IL$^{(0,\alpha)}$PE family was introduced, we represent buoyancy in both the PE and EB reductions by the centered polynomial expansion
\begin{equation}
  \vartheta(\bx,z,t)
  =
  \bar\vartheta(\bx,t)
  +
  \sum_{n=1}^{\alpha}
  \left(\sigma^n-\overline{\sigma^n}\right)
  \vartheta_{\sigma^n}(\bx,t),
  \label{eq:buoyancy-polynomial}
\end{equation}
where
\begin{equation}
  \overline{\sigma^n}
  :=
  \frac12\int_{-1}^{1}\sigma^n\,\dd\sigma
  =
  \begin{cases}
    0, & n\text{ odd},\\
    (n+1)^{-1}, & n\text{ even}.
  \end{cases}
  \label{eq:sigma-moments}
\end{equation}
The centering ensures that $\bar\vartheta$ is the vertical mean buoyancy. The specializations $\alpha=0$ and $\vartheta=g'$ recover IL$^0$ and HL, respectively.

Since $D_t\sigma=0$, substitution of \eqref{eq:buoyancy-polynomial} into $D_t\vartheta=0$ gives
\begin{equation}
  \bar D_t\bar\vartheta=0,
  \qquad
  \bar D_t\vartheta_{\sigma^n}=0,
  \quad n=1,\ldots,\alpha.
  \label{eq:buoyancy-coefficient-advection}
\end{equation}
Together with \eqref{eq:reduced-continuity}, these equations imply advection of $N=\alpha+2$ areal densities, corresponding to the general collection introduced in Section~\ref{sec:EP}:
\begin{equation}
  a_1=h\,\dd^2\bx,
  \qquad
  a_2=h\bar\vartheta\,\dd^2\bx,
  \qquad
  a_{n+2}=h\vartheta_{\sigma^n}\,\dd^2\bx,
  \quad n=1,\ldots,\alpha.
  \label{eq:reduced-advected-densities}
\end{equation}
They satisfy
\begin{equation}
  (\partial_t+\pounds_{\bar u})a_i=0,
  \qquad i=1,\ldots,N,
  \label{eq:reduced-density-advection}
\end{equation}
and their constrained variations are
\begin{equation}
  \delta a_i=-\pounds_{\bar\xi}a_i,
  \qquad i=1,\ldots,N.
  \label{eq:reduced-density-variations}
\end{equation}
In particular, $\delta h=-\nabla\cdot h\bar{\bxi}$, while each buoyancy coefficient varies as a scalar, $\delta\vartheta_{\sigma^n}=-\bar{\bxi}\cdot\nabla\vartheta_{\sigma^n}$, and similarly for $\bar\vartheta$.

\subsection{The reduced Lagrangian family}

Substituting the columnar velocity \eqref{eq:columnar-horizontal-velocity}--\eqref{eq:columnar-vertical-velocity} and the buoyancy representation \eqref{eq:buoyancy-polynomial} into the parent Lagrangian \eqref{eq:parent-Lagrangian}, with $D=1$, and integrating over depth gives the reduced Lagrangian. The horizontal kinetic energy and Coriolis contributions are
\begin{equation}
  \int_{-h}^0\frac12|\bar{\bu}|^2+\bar{\bu}\cdot\mathbf f\,\dd z
  =
  \frac12h|\bar{\bu}|^2+h\bar{\bu}\cdot\mathbf f.
  \label{eq:reduced-horizontal-kinetic}
\end{equation}
The vertical kinetic energy, retained only in the EB branch, is
\begin{equation}
  \int_{-h}^0\frac12w^2\,\dd z
  =
  \frac12(\nabla\cdot\bar{\bu})^2\int_{-h}^0z^2\,\dd z
  =
  \frac16h^3(\nabla\cdot\bar{\bu})^2.
  \label{eq:reduced-vertical-kinetic}
\end{equation}

For the buoyancy potential energy, define the effective buoyancy
\begin{equation}
  \theta
  :=
  \frac12\int_{-1}^{1}(1-\sigma)\vartheta\,\dd\sigma
  =
  -\frac{2}{h^2}\int_{-h}^0z\vartheta\,\dd z.
  \label{eq:effective-buoyancy}
\end{equation}
For the homogeneous-layer models, $\theta=g'$, while for IL$^0$PE and IL$^0$EB, $\theta=\bar\vartheta$. For the polynomial buoyancy representation \eqref{eq:buoyancy-polynomial}, using $z=\frac12h(\sigma-1)$ and \eqref{eq:sigma-moments}, we obtain
\begin{equation}
  \theta
  =
  \bar\vartheta
  -
  \sum_{n=1}^{\alpha}
  \overline{\sigma^{n+1}}\vartheta_{\sigma^n},
  \qquad \alpha\ge1.
  \label{eq:effective-buoyancy-polynomial}
\end{equation}
Thus $\theta$ is determined by the vertically averaged buoyancy and the coefficients of its vertical expansion.

Combining these contributions, the reduced Lagrangian for both branches is
\begin{equation}
  \ell_\varepsilon(\bar u,a)
  =
  \int_{\mathcal D}
    \frac12h|\bar{\bu}|^2
    +h\bar{\bu}\cdot\mathbf f
    +\frac{\varepsilon}{6}h^3(\nabla\cdot\bar{\bu})^2
    -\frac12h^2\theta
    \,\dd^2\bx,
  \label{eq:reduced-Lagrangian}
\end{equation}
where $\varepsilon=0$ gives the hydrostatic PE reductions and $\varepsilon=1$ the nonhydrostatic EB reductions. The advected quantities $a=(a_1,\ldots,a_N)$ are those defined in \eqref{eq:reduced-advected-densities}, with $N=1$ for HL, $N=2$ for IL$^0$, and $N=\alpha+2$ for IL$^{(0,\alpha)}$.

The reduced Lagrangian \eqref{eq:reduced-Lagrangian}, together with the constrained variations \eqref{eq:reduced-velocity-variation} and \eqref{eq:reduced-density-variations}, defines an Euler--Poincar\'e variational principle on $\X(\mathcal D)\times V(\mathcal D)^*$ for each member of the PE and EB families. The common geometric structure is inherited from the parent models through the Lie-algebra-structure-preserving columnar reduction. The distinction between the two branches is entirely contained in the vertical kinetic-energy contribution proportional to $\varepsilon$.

\subsection{Momentum and generalized potential vorticity}

The momentum one-form density associated with the reduced Lagrangian \eqref{eq:reduced-Lagrangian} is $m_\varepsilon=\delta\ell_\varepsilon/\delta\bar u$, with Cartesian coefficients
\begin{equation}
  \mathbf m_\varepsilon
  =
  h(\bar{\bu}+\mathbf f)
  -
  \frac{\varepsilon}{3}\nabla\left(h^3\nabla\cdot\bar{\bu}\right).
  \label{eq:reduced-momentum}
\end{equation}
The second term follows by integrating the variation of the vertical kinetic energy by parts. The specific momentum and its associated generalized potential vorticity are defined by
\begin{equation}
  \mathbf v_\varepsilon
  :=
  \frac{\mathbf m_\varepsilon}{h},
  \qquad
  q_\varepsilon
  :=
  \frac1h\hat{\mathbf z}\cdot\nabla\times\mathbf v_\varepsilon.
  \label{eq:reduced-PV}
\end{equation}
For $\varepsilon=0$, $\mathbf v_0=\bar{\bu}+\mathbf f$ and $q_0$ is the usual shallow-water potential vorticity. For $\varepsilon=1$, the nonhydrostatic contribution modifies the specific momentum and its curl.

Write the advected densities as $a_i=A_i\,\dd^2\bx$, $i=1,\ldots,N$, with $A_1=h$ and $A_i/h$ equal to the corresponding advected buoyancy coefficient for $i\ge2$, as defined in \eqref{eq:reduced-advected-densities}. Let $b_i=\delta\ell_\varepsilon/\delta a_i$. The Euler--Poincar\'e equation \eqref{eq:general-EP} becomes
\begin{equation}
  (\partial_t+\pounds_{\bar u})m_\varepsilon
  =
  \sum_{i=1}^{N}b_i\diamond a_i
  =
  \sum_{i=1}^{N}A_i\,\dd b_i\otimes\dd^2\bx.
  \label{eq:reduced-EP-momentum}
\end{equation}
Dividing by the advected thickness density $a_1=h\,\dd^2\bx$ gives the equation for the specific momentum one-form:
\begin{equation}
  (\partial_t+\pounds_{\bar u})(\mathbf v_\varepsilon\cdot\dd\bx)
  =
  \sum_{i=1}^{N}\frac{A_i}{h}\,\dd b_i.
  \label{eq:specific-momentum-evolution}
\end{equation}

To evaluate the right-hand side, note that $\theta$ is a linear combination of the advected buoyancy coefficients, with constant coefficients determined by \eqref{eq:effective-buoyancy-polynomial}. Hence the reduced potential energy is $-h^2\theta/2$. For $i\ge2$, its functional derivatives satisfy
\begin{equation}
  b_i=-\frac12h\,\frac{\partial\theta}{\partial(A_i/h)}.
  \label{eq:buoyancy-functional-derivatives}
\end{equation}
Since the coefficients $\partial\theta/\partial(A_i/h)$ are constants, it follows that
\begin{equation}
  \sum_{i=2}^{N}\frac{A_i}{h}\,\dd b_i
  =
  -\frac12\theta\,\dd h.
  \label{eq:buoyancy-diamond-sum}
\end{equation}
The remaining term, $\dd b_1$, is exact. Therefore, the specific momentum equation may be written
\begin{equation}
  (\partial_t+\pounds_{\bar u})(\mathbf v_\varepsilon\cdot\dd\bx)
  =
  \dd b_1-\frac12\theta\,\dd h.
  \label{eq:specific-momentum-theta}
\end{equation}

The Euler--Poincar\'e momentum equation can equivalently be expressed in terms of the horizontal acceleration. Substituting the specific momentum \eqref{eq:reduced-PV} into \eqref{eq:specific-momentum-theta} and using the continuity equation \eqref{eq:reduced-continuity}, one obtains, after some algebra,
\begin{equation}
  \bar D_t\bar{\bu}
  +f\hat{\mathbf z}\times\bar{\bu}
  =
  -\theta\nabla h
  -\frac12h\nabla\theta
  +\varepsilon\mathbf F_{\mathrm{NH}},
  \label{eq:reduced-acceleration}
\end{equation}
where the (nonhydrostatic) Green--Naghdi force is
\begin{equation}
  \mathbf F_{\mathrm{NH}}
  :=
  \frac{1}{3h}\nabla\left(
    h^3\left(
      \bar D_t(\nabla\cdot\bar{\bu})
      -(\nabla\cdot\bar{\bu})^2
    \right)
  \right).
  \label{eq:nonhydrostatic-force}
\end{equation}
Equivalently, using $\bar D_t h=-h\nabla\cdot\bar{\bu}$,
\begin{equation}
  \mathbf F_{\mathrm{NH}}
  =
  -\frac{1}{3h}\nabla\left(h^2\bar D_t^2h\right).
  \label{eq:nonhydrostatic-force-thickness}
\end{equation}
Setting $\varepsilon=0$ recovers the hydrostatic PE family, while $\varepsilon=1$ gives the corresponding nonhydrostatic EB family. The buoyancy force has the same form in both branches, with $\theta=g'$ for HL, $\theta=\bar\vartheta$ for IL$^0$, and $\theta$ given by \eqref{eq:effective-buoyancy-polynomial} for IL$^{(0,\alpha)}$.

Let $\gamma_t$ be a closed material curve transported by $\bar{\bu}$. Integrating \eqref{eq:specific-momentum-theta} along $\gamma_t$ and applying the Lie derivative theorem yields the Kelvin--Noether circulation law
\begin{equation}
  \frac{\dd}{\dd t}
  \oint_{\gamma_t}\mathbf v_\varepsilon\cdot\dd\bx
  =
  -\frac12\oint_{\gamma_t}\theta\,\dd h
  =
  \frac12\oint_{\gamma_t}h\,\dd\theta.
  \label{eq:reduced-Kelvin-Noether}
\end{equation}
The second equality follows because the integral of $\dd(h\theta)$ around a closed curve vanishes.

Taking the exterior derivative of \eqref{eq:specific-momentum-theta} and using the commutativity of $\dd$ and $\pounds_{\bar u}$ gives
\begin{equation}
  (\partial_t+\pounds_{\bar u})(hq_\varepsilon\,\dd^2\bx)
  =
  -\frac12\dd\theta\wedge\dd h
  =
  \frac12\dd h\wedge\dd\theta.
  \label{eq:reduced-vorticity-evolution}
\end{equation}
Since $h\,\dd^2\bx$ is advected, the corresponding potential-vorticity equation is
\begin{equation}
  \bar D_t q_\varepsilon
  =
  \frac12\hat{\mathbf z}\cdot
  \nabla\log h\times\nabla\theta.
  \label{eq:reduced-PV-evolution}
\end{equation}

For HLPE and HLEB, $\theta=g'$ is constant, so $q_\varepsilon$ is materially conserved. For IL$^0$PE and IL$^0$EB, $\theta=\bar\vartheta$, giving the familiar baroclinic PV source. For IL$^{(0,\alpha)}$PE and IL$^{(0,\alpha)}$EB, $\theta$ is the combination of advected buoyancy coefficients given by \eqref{eq:effective-buoyancy-polynomial}. Thus the PV evolution law has the same form throughout both model hierarchies; the distinction between PE and EB resides in the generalized momentum and potential vorticity.

\section{A common construction of particle-relabeling generators}
\label{sec:common-Noether}

We build on the Eulerian particle-relabeling framework of \cite{Cotter-Holm-12}, who derive local conservation laws and conserved integrals from transported symmetry generators that preserve the advected quantities. Here we construct generators for the PE and EB parent models and their columnar reductions and explicitly evaluate their Noether pairings as vorticity-dependent conserved integrals. The construction uses a transported differential form to define a displacement that preserves the fluid volume or layer-thickness measure and satisfies the relabeling transport equation. The generators belong to the admissible class specified in Section~\ref{sec:EP}, and preservation of the Lagrangian is checked separately for each model. For the stratified columnar reductions, we use the weaker condition of preserving the effective buoyancy entering the Lagrangian, without requiring preservation of every advected buoyancy coefficient.

Let $\mu=\rho\,\dd^nx$ be a positive density on $M$ satisfying
\begin{equation}
  (\partial_t+\pounds_u)\mu
  =
  0.
  \label{eq:common-measure-transport}
\end{equation}
In the applications below, $\mu$ is the fluid volume measure or the layer-thickness measure. Let $\beta$ be a differential form of degree $n-1$, hence a one-form in two dimensions or a two-form in three dimensions, satisfying
\begin{equation}
  \dd\beta
  =
  0,
  \qquad
  (\partial_t+\pounds_u)\beta
  =
  0.
  \label{eq:common-form-conditions}
\end{equation}
Define the displacement $\xi=\boldsymbol\xi\cdot\nabla$ by
\begin{equation}
  \iota_\xi\mu
  =
  \beta.
  \label{eq:common-geometric-generator}
\end{equation}
Positivity of $\rho$ ensures that this relation determines $\xi$ uniquely.

The condition $\dd\beta=0$ implies preservation of the measure. Indeed, Cartan's formula gives
\begin{equation}
  \pounds_\xi\mu
  =
  \dd\iota_\xi\mu
  =
  \dd\beta
  =
  0,
  \label{eq:common-measure-preservation}
\end{equation}
or, in Cartesian coordinates, $\nabla\cdot\rho\boldsymbol\xi=0$. Transport of $\xi$ follows by applying $\partial_t+\pounds_u$ to \eqref{eq:common-geometric-generator}. Using the identity $\pounds_u\iota_\xi-\iota_\xi\pounds_u=\iota_{[u,\xi]}$, together with the transport equations for $\mu$ and $\beta$, we obtain
\begin{equation}
  \iota_{\partial_t\xi+[u,\xi]}\mu
  =
  0.
\end{equation}
Since $\rho>0$, it follows that
\begin{equation}
  \partial_t\xi+[u,\xi]
  =
  0.
  \label{eq:common-geometric-transport}
\end{equation}
Thus the displacement satisfies the relabeling transport equation and has vanishing constrained velocity variation.

The forms $\beta$ used below are constructed from materially conserved scalars, which may be advected scalar fields, functions or ratios of the advected quantities, or quantities whose material conservation follows from the momentum equation. In two and three dimensions, respectively, we use
\begin{equation}
  \beta=\dd G(\chi),\qquad \beta=\dd\chi\wedge\dd G(\chi,\tilde\chi),
  \label{eq:common-scalar-construction}
\end{equation}
where $\chi$ and $\tilde\chi$ are materially conserved scalars and $G$ is a smooth function of the indicated arguments. These forms satisfy \eqref{eq:common-form-conditions}: they are closed because $\dd^2=0$ and transported because exterior differentiation commutes with $\partial_t+\pounds_u$, which obeys the product rule for exterior products. The defining scalars need not coincide with the advected variables entering the Lagrangian.

The construction therefore provides a transported generator preserving $\mu$. With the reference fields held fixed, it induces $\delta_\xi a=-\pounds_\xi a$. Depending on the chosen form $\beta$ and the advected quantities, these variations may all vanish, \emph{but this must be checked separately}. Preservation of every advected quantity is sufficient for Lagrangian preservation, \eqref{eq:Lagrangian-preservation}; more generally, their combined contribution to $\delta_\xi\ell$ may vanish even when individual fields change.

When $\delta_\xi\ell=0$, the Noether quantity $J_\xi$ defined in \eqref{eq:Noether-pairing} is conserved. The following sections specify the transported measure and form, verify the remaining symmetry conditions, and evaluate the pairing using the common assumption that spatial boundary terms vanish. The equations for the transported form ensure measure preservation and generator transport; membership in the admissible class remains an assumption and is not a consequence of those equations alone.

\section{Explicit Noether quantities in the parent models}
\label{sec:parent-Noether}

\subsection{Specific momentum and Ertel potential vorticity}

For the parent models, let $u_3=\bu\cdot\nabla+w\partial_z$ denote the full velocity vector field. The specific momentum one-form associated with \eqref{eq:parent-Lagrangian} is
\begin{equation}
  V_\varepsilon
  :=
  (\bu+\mathbf f)\cdot\dd\bx+\varepsilon w\,\dd z,
  \qquad
  \varepsilon\in\{0,1\}.
  \label{eq:parent-specific-momentum}
\end{equation}
Indeed, the momentum one-form density obtained from \eqref{eq:parent-functional-derivatives} is $V_\varepsilon\otimes(D\,\dd^3x)$. Thus IL$^\infty$PE has $V_0=(\bu+\mathbf f)\cdot\dd\bx$, whereas IL$^\infty$EB has $V_1=(\bu+\mathbf f)\cdot\dd\bx+w\,\dd z$. The absence of a vertical momentum component in the PE case does not imply that $w$ vanishes: both models transport fluid parcels with the full velocity $u_3$.

The preserved measure in the common construction is the fluid volume form. On imposing $D=1$, we have
\begin{equation}
  \mu=\dd^3x=\dd^2\bx\,\dd z,
  \qquad
  (\partial_t+\pounds_{u_3})\mu=0,
  \label{eq:parent-volume-transport}
\end{equation}
by incompressibility in \eqref{eq:parent-kinematics}. This transport law holds on the material domain $M_t$; the interface kinematic condition ensures that its lower boundary moves with the fluid. The measure $\mu$ represents volume rather than mass, consistently with the Boussinesq constraint $D=1$ and the normalization of the parent Lagrangian by $\rho_{\mathrm{ref}}$.

Using the volume-density and scalar diamond contributions in \eqref{eq:general-EP}, the parent functional derivatives give, at $D=1$,
\begin{equation}
  (\partial_t+\pounds_{u_3})V_\varepsilon
  =
  \dd B_\varepsilon-z\,\dd\vartheta,
  \qquad
  B_\varepsilon
  :=
  \frac12|\bu|^2+\frac{\varepsilon}{2}w^2
  +\bu\cdot\mathbf f+\vartheta z-p.
  \label{eq:parent-specific-momentum-evolution}
\end{equation}
Here $B_\varepsilon=\delta\ell_\varepsilon/\delta D$ evaluated at $D=1$. The scalar diamond contribution is $-z\,\dd\vartheta$; together with the term $\dd(\vartheta z)$ contained in $\dd B_\varepsilon$, it gives the buoyancy force $\vartheta\,\dd z$.

Define the Ertel potential vorticity $Q_\varepsilon$ by
\begin{equation}
  \dd V_\varepsilon\wedge\dd\vartheta
  =
  Q_\varepsilon\,\mu.
  \label{eq:parent-Ertel-definition}
\end{equation}
We use an uppercase $Q_\varepsilon$ to distinguish the parent Ertel potential vorticity from the reduced generalized potential vorticity $q_\varepsilon$ in \eqref{eq:reduced-PV}. With the orientation $\dd^3x=\dd x^1\wedge\dd x^2\wedge\dd z$, \eqref{eq:parent-Ertel-definition} reads
\begin{equation}
  Q_\varepsilon
  =
  \left(
    \nabla_3\times(\bu+\mathbf f,\varepsilon w)
  \right)\cdot\nabla_3\vartheta.
  \label{eq:parent-Ertel-coordinate}
\end{equation}
The curl here is three-dimensional. In particular, the hydrostatic expression retains the contributions from vertical shear of the horizontal velocity.

To establish material conservation, take the exterior derivative of \eqref{eq:parent-specific-momentum-evolution}:
\begin{equation}
  (\partial_t+\pounds_{u_3})\dd V_\varepsilon
  =
  -\dd z\wedge\dd\vartheta.
  \label{eq:parent-vorticity-form-evolution}
\end{equation}
Buoyancy advection implies $(\partial_t+\pounds_{u_3})\dd\vartheta=0$. Consequently,
\begin{equation}
  (\partial_t+\pounds_{u_3})
  \left(\dd V_\varepsilon\wedge\dd\vartheta\right)
  =
  -\dd z\wedge\dd\vartheta\wedge\dd\vartheta
  =
  0.
  \label{eq:parent-Ertel-form-transport}
\end{equation}
Combining this identity with \eqref{eq:parent-volume-transport} and \eqref{eq:parent-Ertel-definition} yields
\begin{equation}
  D_tQ_\varepsilon=0.
  \label{eq:parent-Ertel-advection}
\end{equation}
Thus $\vartheta$ and $Q_\varepsilon$ are materially conserved scalars for both parents, while $\mu$ is their common transported volume form. These are the ingredients used below to construct the relabeling generator.

\subsection{Construction of the generator}

Let $G=G(\vartheta,Q_\varepsilon)$ be a smooth function. We apply the common construction of Section~\ref{sec:common-Noether} with the volume form $\mu=\dd^3x$ and the two-form
\begin{equation}
  \beta_\varepsilon
  :=
  \dd\vartheta\wedge\dd G(\vartheta,Q_\varepsilon),
  \qquad
  \iota_{\xi_3}\mu=\beta_\varepsilon.
  \label{eq:parent-generator-definition}
\end{equation}
Both $\vartheta$ and $Q_\varepsilon$ are materially conserved, so $\beta_\varepsilon$ is closed and transported. Consequently, the resulting displacement preserves volume and satisfies the relabeling transport equation:
\begin{equation}
  \pounds_{\xi_3}\mu=0,
  \qquad
  \partial_t\xi_3+[u_3,\xi_3]=0.
  \label{eq:parent-generator-transport}
\end{equation}
In coordinates, $\xi_3=\boldsymbol\xi_3\cdot\nabla_3$ has coefficients
\begin{equation}
  \boldsymbol\xi_3
  =
  \nabla_3\vartheta\times\nabla_3G.
  \label{eq:parent-generator-coordinate}
\end{equation}

The displacement also preserves buoyancy, since
\begin{equation}
  \pounds_{\xi_3}\vartheta
  =
  \boldsymbol\xi_3\cdot\nabla_3\vartheta
  =
  0.
  \label{eq:parent-generator-buoyancy}
\end{equation}
Thus, at $D=1$, the constrained variations satisfy $\delta_\xi D=0$, $\delta_\xi\vartheta=0$, and $\delta_\xi u_3=0$. With the prescribed rotation potential and pressure multiplier held fixed, these conditions preserve the parent Lagrangian \eqref{eq:parent-Lagrangian} for both $\varepsilon=0$ and $\varepsilon=1$. This construction therefore satisfies the stronger symmetry condition \eqref{eq:individual-preservation}. Although \eqref{eq:parent-generator-coordinate} also gives $\pounds_{\xi_3}Q_\varepsilon=0$, this is not an additional requirement for Lagrangian preservation: $Q_\varepsilon$ is derived from the momentum and buoyancy fields.

\subsection{Evaluation of the Noether pairing}

At $D=1$, the momentum one-form density is
\begin{equation}
    m_\varepsilon = \frac{\delta\ell_\varepsilon}{\delta u_3} = V_\varepsilon\otimes\mu, 
\end{equation}
with $\mu=\dd^3x$. Since $V_\varepsilon\wedge\mu$ is a four-form on a three-dimensional domain, it vanishes, and the contraction product rule gives $(\iota_{\xi_3}V_\varepsilon)\mu=V_\varepsilon\wedge\iota_{\xi_3}\mu$. Recalling that $G=G(\vartheta,Q_\varepsilon)$ enters through $\iota_{\xi_3}\mu=\dd\vartheta\wedge\dd G$ in \eqref{eq:parent-generator-definition}, the Noether pairing is
\begin{equation}
  J_{\varepsilon,\xi_3}
  =
  \left\langle
    m_\varepsilon,\xi_3
  \right\rangle
  =
  \int_{M_t}(\iota_{\xi_3}V_\varepsilon)\,\mu
  =
  \int_{M_t}V_\varepsilon\wedge\iota_{\xi_3}\mu
  =
  \int_{M_t}V_\varepsilon\wedge\dd\vartheta\wedge\dd G.
  \label{eq:parent-Noether-form-pairing}
\end{equation}

To evaluate \eqref{eq:parent-Noether-form-pairing}, observe that
\begin{equation}
  \dd\left(G(\vartheta,Q_\varepsilon)V_\varepsilon\wedge\dd\vartheta\right)
  =
  V_\varepsilon\wedge\dd\vartheta\wedge\dd G(\vartheta,Q_\varepsilon)
  +
  G(\vartheta,Q_\varepsilon)\,\dd V_\varepsilon\wedge\dd\vartheta.
\end{equation}
Using \eqref{eq:parent-Ertel-definition} and omitting the integral of the exact form under our spatial boundary convention gives
\begin{equation}
  J_{\varepsilon,\xi_3}=-\int_{M_t}Q_\varepsilon G(\vartheta,Q_\varepsilon)\,\dd^3x=:\mathcal C_{\varepsilon,G}.
  \label{eq:parent-Noether-evaluated}
\end{equation}
Thus the vorticity-dependent integral produced by the generator is itself the conserved pairing. Independently, buoyancy and Ertel-PV advection imply conservation of
\begin{equation}
  \mathcal C_{\varepsilon,\Phi}=\int_{M_t}\Phi(\vartheta,Q_\varepsilon)\,\dd^3x
  \label{eq:parent-Ertel-family}
\end{equation}
for every smooth $\Phi$ for which the integral exists. The relation between the generated and general families is
\begin{equation}
  G(\vartheta,Q)=-\int_0^1\partial_2\Phi(\vartheta,\tau Q)\,\dd\tau,\qquad -QG(\vartheta,Q)=\Phi(\vartheta,Q)-\Phi(\vartheta,0).
  \label{eq:parent-Phi-decomposition}
\end{equation}
This definition is smooth at $Q=0$. The buoyancy-only integral $\int_{M_t}\Phi(\vartheta,0)\,\dd^3x$ is conserved by advection. It belongs to the general family independently of the vorticity-dependent pairing and need not be added to define $\mathcal C_{\varepsilon,G}$.

\subsection{The Euler--Boussinesq bracket and its Casimirs}
\label{sec:parent-Hamiltonian}

For IL$^\infty$EB, the velocity Legendre transform is regular at positive volume density $D$, with $\mathbf m=D(\bu+\mathbf f,w)$. The resulting Hamiltonian is the standard Euler--Boussinesq Hamiltonian \cite{Holm-etal-84}, with incompressibility imposed by the pressure multiplier. With $a=(D\,\dd^3x,\vartheta)$, the Lie--Poisson bracket \eqref{eq:general-LP-bracket} gives the fluid specialization of the bracket introduced by \cite{Morrison-Greene-80} for magnetohydrodynamics. Its Ertel Casimirs \cite{Holm-etal-84} are
\begin{equation}
  \mathcal C_\Phi=\int_{M_t}D\Phi(\vartheta,Q_D)\,\dd^3x,\qquad Q_D=\frac1D\left(\nabla_3\times\frac{\mathbf m}{D}\right)\cdot\nabla_3\vartheta.
  \label{eq:parent-density-Casimir}
\end{equation}
At $D=1$, this family reduces to \eqref{eq:parent-Ertel-family} with $\varepsilon=1$ and includes the buoyancy-only integrals. These statements use the common assumption that spatial boundary terms vanish.

For IL$^\infty$PE, the momentum has no vertical component even though parcels have a vertical velocity. The full velocity Legendre transform is consequently degenerate, and the EB Hamiltonian cannot simply be specialized by setting $\varepsilon=0$. The generator and evaluated pairing above remain valid, as do the Ertel and buoyancy-only advection integrals. We make no Casimir claim for this parent based on a regular Legendre transform.

\section{Explicit Noether quantities in the reduced models}
\label{sec:reduced-Noether}

\subsection{The reduced generator and endpoint pairing}

Recall $\mathbf v_\varepsilon=\mathbf m_\varepsilon/h$ and $hq_\varepsilon=\hat{\mathbf z}\cdot\nabla\times\mathbf v_\varepsilon$, with $h>0$. Let $\chi$ be materially conserved by $\bar u$ and let $G$ be smooth. Apply Section~\ref{sec:common-Noether} with the transported measure $h\,\dd^2\bx$ and the closed transported one-form $\dd G(\chi)$. The displacement is defined by
\begin{equation}
  \iota_\xi(h\,\dd^2\bx)=\dd G(\chi),\qquad \bxi=-\frac1h\hat{\mathbf z}\times\nabla G(\chi).
  \label{eq:reduced-geometric-generator}
\end{equation}
The construction gives
\begin{equation}
  \partial_t\xi+[\bar u,\xi]=0,\qquad \nabla\cdot h\bxi=0,\qquad \bxi\cdot\nabla\chi=0.
  \label{eq:reduced-generator-properties}
\end{equation}
Thus the displacement preserves velocity, thickness measure, and its defining scalar. Writing $V_\varepsilon=\mathbf v_\varepsilon\cdot\dd\bx$, the pairing is evaluated using
\begin{equation}
  J_{\varepsilon,\xi}=\int_{\mathcal D}V_\varepsilon\wedge\dd G(\chi),\qquad V_\varepsilon\wedge\dd G(\chi)=G(\chi)\,\dd V_\varepsilon-\dd\bigl(G(\chi)V_\varepsilon\bigr).
  \label{eq:reduced-endpoint-pairing}
\end{equation}
Under the spatial boundary convention of Section~\ref{sec:EP}, this gives
\begin{equation}
  J_{\varepsilon,\xi}=\int_{\mathcal D}hq_\varepsilon G(\chi)\,\dd^2\bx.
  \label{eq:reduced-evaluated-pairing}
\end{equation}
Lagrangian preservation remains to be verified in each application.

The reduced bracket is the advected-density specialization of \eqref{eq:general-LP-bracket}. Using the density coefficients $A_i$ introduced above, and writing $\delta\mathcal F/\delta m_\varepsilon=\mathbf F_m\cdot\nabla$ and $\mathcal F_i=\delta\mathcal F/\delta A_i$, its coordinate form is
\begin{equation}
  \{\mathcal F,\mathcal K\}_N
  =
  -\int_{\mathcal D}\mathbf m_\varepsilon\cdot[\mathbf F_m,\mathbf K_m]\,\dd^2\bx
  +
  \sum_{i=1}^{N}\int_{\mathcal D}A_i(\mathbf K_m\cdot\nabla\mathcal F_i-\mathbf F_m\cdot\nabla\mathcal K_i)\,\dd^2\bx.
  \label{eq:reduced-LP-bracket}
\end{equation}
The commutator denotes the coordinate coefficients of \eqref{eq:vector-bracket}. For the PE hierarchy this bracket is given by \cite{Beron-21-POFb}. For the EB hierarchy, including the IL$^{(0,\alpha)}$EB models introduced here, it follows by the same semidirect-product reduction: the velocity Lie algebra and the advected densities are unchanged, while the vertical kinetic energy changes the momentum--velocity relation and the Hamiltonian. The Hamiltonian follows from the Legendre transform of \eqref{eq:reduced-Lagrangian}:
\begin{equation}
  \mathcal H_\varepsilon(m_\varepsilon,a)
  =
  \langle m_\varepsilon,\bar u\rangle-\ell_\varepsilon(\bar u,a)
  =
  \int_{\mathcal D}\frac12h|\bar{\bu}|^2+\frac{\varepsilon}{6}h^3(\nabla\cdot\bar{\bu})^2+\frac12h^2\theta\,\dd^2\bx,
  \label{eq:reduced-Hamiltonian}
\end{equation}
where $\bar u$ is expressed in terms of $m_\varepsilon$ using \eqref{eq:reduced-momentum}. We assume positive thickness and function spaces on which this momentum relation is invertible. The bracket and Casimir statements below use the common convention that spatial boundary terms vanish.

\subsection{Stratified layers: IL$^{(0,\alpha)}$PE and IL$^{(0,\alpha)}$EB}
\label{sec:stratified-Noether}

For $\alpha\ge1$, the effective buoyancy $\theta=\bar\vartheta-\sum_{n=1}^{\alpha}\overline{\sigma^{n+1}}\vartheta_{\sigma^n}$ is a linear combination of advected buoyancy coefficients with constant weights and is therefore materially conserved. Taking $\chi=\theta$ gives
\begin{equation}
  \iota_\xi(h\,\dd^2\bx)=\dd G(\theta),\qquad \bxi=-\frac1h\hat{\mathbf z}\times\nabla G(\theta),\qquad \delta_\xi\bar u=0,\quad\delta_\xi h=0,\quad\delta_\xi\theta=0.
  \label{eq:stratified-generator}
\end{equation}
Since \eqref{eq:reduced-Lagrangian} depends on the buoyancy coefficients only through $\theta$, these conditions imply $\delta_\xi\ell_\varepsilon=0$. The individual coefficients need not be preserved: their variations need only satisfy $\delta_\xi\bar\vartheta-\sum_{n=1}^{\alpha}\overline{\sigma^{n+1}}\delta_\xi\vartheta_{\sigma^n}=0$. The conserved pairing is
\begin{equation}
  J_{\varepsilon,\xi}=\mathcal C_{\varepsilon,G},\qquad \mathcal C_{\varepsilon,G}=\int_{\mathcal D}hq_\varepsilon G(\theta)\,\dd^2\bx.
  \label{eq:stratified-Noether-pairing}
\end{equation}
This does not require material conservation of $q_\varepsilon$, which generally has the baroclinic source \eqref{eq:reduced-PV-evolution}. A separate conserved family follows directly from joint advection of the buoyancy coefficients and transport of the thickness measure. For any smooth function $F$ in the admissible class,
\begin{equation}
  \mathcal A_F=\int_{\mathcal D}hF(\bar\vartheta,\vartheta_\sigma,\ldots,\vartheta_{\sigma^\alpha})\,\dd^2\bx,\qquad \frac{\dd\mathcal A_F}{\dd t}=0.
  \label{eq:stratified-advective-integrals}
\end{equation}

The full bracket is \eqref{eq:reduced-LP-bracket} with $N=\alpha+2$, $A_1=h$, $A_2=h\bar\vartheta$, and $A_{n+2}=h\vartheta_{\sigma^n}$, $n=1,\ldots,\alpha$. For the PE models, the Casimir family identified in \cite{Beron-21-POFb} can be written as
\begin{equation}
  \mathcal C_{\varepsilon,c,F}=c\int_{\mathcal D}hq_\varepsilon\,\dd^2\bx+\mathcal A_F,
  \label{eq:stratified-Casimir-family}
\end{equation}
where $c$ is constant. The same family applies to the EB models because the bracket in momentum and density variables is unchanged. Thus the joint buoyancy integrals $\mathcal A_F$ are Casimirs, and a constant multiple of the total generalized circulation may be added.

In contrast, the vorticity-dependent Noether integrals $\mathcal C_{\varepsilon,G}$ are \emph{generally not Casimirs} for nonconstant $G$. Their momentum derivative generates a displacement along contours of $\theta$, which need not preserve every independent buoyancy coefficient. Their conservation therefore depends on the particular Lagrangian through its dependence on $\theta$. If only the projected densities $h$ and $h\theta$ are retained, the same integrals are Casimirs of that smaller bracket; this does not make them Casimirs of the full stratified bracket.

\subsection{Thermal layers: IL$^0$PE and IL$^0$EB}
\label{sec:thermal-Noether}

For thermal layers, $\theta=\bar\vartheta$, and the generator specializes to
\begin{equation}
  \iota_\xi(h\,\dd^2\bx)=\dd G(\bar\vartheta),\qquad \bxi=-\frac1h\hat{\mathbf z}\times\nabla G(\bar\vartheta).
  \label{eq:thermal-generator}
\end{equation}
Preservation of thickness and effective buoyancy now preserves \emph{every advected quantity}, $h\,\dd^2\bx$ and $h\bar\vartheta\,\dd^2\bx$. The generator therefore satisfies the stronger symmetry condition \eqref{eq:individual-preservation}, and the conserved pairing is
\begin{equation}
  J_{\varepsilon,\xi}=\mathcal C_{\varepsilon,G},\qquad \mathcal C_{\varepsilon,G}=\int_{\mathcal D}hq_\varepsilon G(\bar\vartheta)\,\dd^2\bx.
  \label{eq:thermal-Noether-pairing}
\end{equation}
Separately, buoyancy advection and thickness transport conserve $\mathcal A_F=\int_{\mathcal D}hF(\bar\vartheta)\,\dd^2\bx$ for any smooth function $F$ in the admissible class.

The thermal case corresponds to $\alpha=0$, and its bracket is \eqref{eq:reduced-LP-bracket} with $N=2$, $A_1=h$, and $A_2=h\bar\vartheta$. The IL$^0$PE Hamiltonian structure is given by \cite{Dellar-03}, and the IL$^0$EB bracket follows by the same Legendre-transform argument used above. For both branches, the Casimir family, also given for thermal Green--Naghdi dynamics by \cite{Holm-Street-26}, is
\begin{equation}
  \mathcal C_{\varepsilon,F,G}=\mathcal C_{\varepsilon,G}+\mathcal A_F.
  \label{eq:thermal-Casimir-family}
\end{equation}
Thus the vorticity-dependent Noether integrals are \emph{Casimirs of the full thermal bracket}, as are the buoyancy-only integrals obtained directly from advection. Unlike the stratified case, the coefficient $G(\bar\vartheta)$ multiplying $q_\varepsilon$ need not be constant.

\subsection{Homogeneous layers: HLPE and HLEB}
\label{sec:homogeneous-Noether}

For homogeneous layers, $\theta=g'$ is constant, so \eqref{eq:reduced-PV-evolution} gives $(\partial_t+\bar{\bu}\cdot\nabla)q_\varepsilon=0$. Taking $\chi=q_\varepsilon$ gives
\begin{equation}
  \iota_\xi(h\,\dd^2\bx)=\dd G(q_\varepsilon),\qquad \bxi=-\frac1h\hat{\mathbf z}\times\nabla G(q_\varepsilon).
  \label{eq:homogeneous-generator}
\end{equation}
The thickness density $h\,\dd^2\bx$ is the only advected quantity entering the Lagrangian. Thus $\delta_\xi\bar u=0$ and $\delta_\xi h=0$ imply $\delta_\xi\ell_\varepsilon=0$, and the conserved pairing is
\begin{equation}
  J_{\varepsilon,\xi}=\mathcal C_{\varepsilon,G},\qquad \mathcal C_{\varepsilon,G}=\int_{\mathcal D}hq_\varepsilon G(q_\varepsilon)\,\dd^2\bx.
  \label{eq:homogeneous-Noether-pairing}
\end{equation}
Writing $\Phi(q)=qG(q)$ identifies this pairing directly with a PV integral. Independently, PV advection and thickness transport conserve the general family
\begin{equation}
  \mathcal C_{\varepsilon,\Phi}=\int_{\mathcal D}h\Phi(q_\varepsilon)\,\dd^2\bx.
  \label{eq:homogeneous-PV-family}
\end{equation}
For a prescribed smooth $\Phi$, the choice $G(q)=\int_0^1\Phi'(\tau q)\,\dd\tau$ is smooth even at $q=0$ and yields $qG(q)=\Phi(q)-\Phi(0)$. Consequently, $\mathcal C_{\varepsilon,\Phi}=\mathcal C_{\varepsilon,G}+\Phi(0)\mathcal V$, where $\mathcal V=\int_{\mathcal D}h\,\dd^2\bx$ is separately conserved. This decomposition does not change the \emph{direct identification} $J_{\varepsilon,\xi}=\mathcal C_{\varepsilon,G}$.

The homogeneous bracket is \eqref{eq:reduced-LP-bracket} with $N=1$ and $A_1=h$. The corresponding Hamiltonian structures and PV Casimirs are discussed by \cite{Holm-etal-84} for shallow-water dynamics and by \cite{Percival-etal-08} for the Green--Naghdi class. Thus \eqref{eq:homogeneous-PV-family}, with arbitrary smooth $\Phi$ in the admissible class, is a Casimir family for both HLPE and HLEB. Its constant-in-PV members are proportional to volume, while its members with $\Phi(0)=0$ admit the direct Noether representation above.

The homogeneous construction uses \emph{a different conserved scalar from the buoyancy-based parent Ertel construction}. Constant buoyancy gives $\dd\vartheta=0$, so the $\vartheta$-Ertel PV and its associated generator vanish. However, the columnar flow materially conserves $z/h$. Extending the reduced specific momentum one-form $V_\varepsilon=\mathbf v_\varepsilon\cdot\dd\bx$ independently of depth gives $\dd V_\varepsilon\wedge\dd(z/h)=q_\varepsilon\,\dd^3x$. For HLPE, this recovers the Ertel interpretation discussed by \cite{Ripa-JFM-95}. The same identity provides a generalized Ertel interpretation for HLEB, although its reduced specific momentum includes the columnar nonhydrostatic correction and differs from the parent one-form containing $w\,\dd z$. The reduced relabeling generator uses the materially conserved scalar $q_\varepsilon$, whose conservation follows from the homogeneous momentum equation.

\section{Summary and Discussion}
\label{sec:discussion}

The construction developed here separates transport of the relabeling generator from preservation of the Lagrangian. A closed transported differential form defines a displacement that preserves the fluid measure and is transported by the flow as a vector field. To generate a conserved Noether pairing, this displacement must also preserve the Lagrangian. In the parent, homogeneous, and thermal models, the generators constructed here preserve every advected quantity. In the stratified columnar models, preservation of thickness and effective buoyancy suffices even when individual buoyancy coefficients change. This weaker symmetry yields \emph{conserved quantities that are generally not Casimirs of the full Lie--Poisson bracket}.

The columnar velocity reduction preserves the Lie-algebra structure, providing a common Euler--Poincar\'e framework for the PE and EB hierarchies. The distinction between the branches enters through their kinetic energies and corresponding momenta. The EB parent retains vertical momentum, while the EB columnar models contain the Green--Naghdi correction to the reduced specific momentum. These differences modify the associated PV without altering the relabeling construction. In momentum and advected-density variables, the reduced Lie--Poisson bracket is the same for both branches; their Hamiltonians follow from the respective Legendre transforms. The same argument supplies the Hamiltonian formulation of the stratified nonhydrostatic columnar models introduced here. For the unrestricted parents, however, the EB velocity Legendre transform is regular whereas the PE transform is degenerate. The hydrostatic-parent conservation results therefore stand independently of a Casimir identification based on a regular Legendre transform.

Three conservation statements have been distinguished:
\begin{enumerate}
    \item conservation of an endpoint pairing by a relabeling symmetry,
    \item conservation of an integral by material advection, and
    \item membership in the kernel of a specified Poisson bracket.
\end{enumerate}
For homogeneous layers, the pairing directly produces thickness-weighted PV integrals whose integrands are the product of PV and a smooth function of PV; the general PV Casimir family additionally includes the independently conserved volume contribution. For thermal layers, the vorticity-dependent Noether integrals and buoyancy-only advection integrals together form the familiar Casimir family. For stratified layers, the joint buoyancy integrals are Casimirs, and a constant multiple of total generalized circulation may be added. The Noether integrals with a nonconstant coefficient depending on effective buoyancy, however, generally are not Casimirs. They become Casimirs if only thickness and effective buoyancy are retained, but this changes the phase space and its bracket. For the EB parent, the Ertel family has the standard Casimir interpretation; for the PE parent, its advection and Noether interpretations remain valid without invoking that Hamiltonian argument.

The homogeneous PV retains an Ertel interpretation even though the buoyancy-based parent Ertel PV vanishes. The columnar flow materially conserves height divided by layer thickness. Using this scalar together with the depth-independent extension of the reduced specific momentum gives the homogeneous PV through the Ertel construction. For HLPE, this recovers the interpretation discussed by \cite{Ripa-JFM-95}; for HLEB, it uses the reduced specific momentum containing the nonhydrostatic correction. This interpretation relies on columnar kinematics: height divided by layer thickness is not generally materially conserved in the unrestricted parents.

\subsection{Spatial boundary terms}

The calculations assume that spatial boundary terms vanish for the fields, generators, and functional variations under consideration. Temporal endpoint terms are retained to obtain the Noether pairing. The cancellation of spatial terms is \emph{not a consequence of impermeability alone}. At a solid lateral boundary, tangency of a reduced generator requires $G(\chi)|_{\partial\mathcal D_j}=c_j$ on each connected component, and the evaluated pairing generally takes the form
\begin{equation}
  J_{\varepsilon,\xi}=\int_{\mathcal D}hq_\varepsilon G(\chi)\,\dd^2\bx-\sum_jc_j\Gamma_{\varepsilon,j},\qquad \Gamma_{\varepsilon,j}=\oint_{\partial\mathcal D_j}\mathbf v_\varepsilon\cdot\dd\bx.
  \label{eq:discussion-boundary-pairing}
\end{equation}
The circulations carry the orientations induced by $\mathcal D$. For a simply connected domain, Stokes' theorem gives $\Gamma_\varepsilon=\int_{\mathcal D}hq_\varepsilon\,\dd^2\bx$. In the homogeneous case, the boundary contribution can therefore be absorbed by replacing $qG(q)$ with $q(G(q)-c)$. The same replacement works in a multiply connected domain when all boundary components share the value $c$. Unequal values generally leave separate circulation contributions.

Such an absorption does not by itself establish differentiability of the resulting functional or admissibility of its momentum derivative. Variation of a homogeneous PV integral produces the boundary term $\oint_{\partial\mathcal D}\Phi'(q_\varepsilon)\delta V_\varepsilon$, whereas variation of a thermal or stratified vorticity integral produces $\oint_{\partial\mathcal D}G(\theta)\delta V_\varepsilon$. Our assumptions require these terms to vanish on the allowed variations. Retaining them requires a specification of the functional class and boundary circulation data. Admissible functionals in fluid Hamiltonian calculations are discussed by \cite{McIntyre-Shepherd-87}, while the dynamic-boundary formulation of \cite{Lewis-etal-86} incorporates boundary geometry into the Hamiltonian description. An analysis of the present solid-wall and material-interface setting along those lines lies beyond the scope of this work.

Periodicity likewise requires consistency of all fields entering the pairing and bracket, rather than periodic velocity and thickness alone. On a $\beta$ plane, an arbitrary function of PV need not be doubly periodic; moreover, nonzero mean planetary vorticity cannot be represented by a globally periodic vector potential. Periodic applications therefore require an appropriate global formulation. For the parent models, horizontal periodicity or decay also leaves contributions at the lid and material interface to be addressed separately.

\subsection{Outlook}

Extension to models with vertical velocity shear requires a reduction compatible with both the velocity Lie algebra and transport of the retained buoyancy variables. Substituting a polynomial velocity ansatz and projecting the equations onto selected vertical modes does not by itself preserve the Hamiltonian structure. A compatible Lie algebra and variational reduction must be established before applying the Euler--Poincar\'e, Legendre-transform, and relabeling arguments developed here. The columnar reduction supplies this compatibility for the present model hierarchies.

\subsection*{Acknowledgments}

The author thanks the Intercultural Outreach Initiative (IOI) in Isabela, Gal\'apagos, for its hospitality during part of this work. This work received no external funding.

\subsection*{Author Contributions}

The sole author carried out all aspects of this work.

\subsection*{Data Availability}

Data sharing is not applicable to this article as no datasets were generated or analyzed during this study.

\section*{Declarations}

\subsection*{Conflict of Interest}

The author declares no conflict of interest.

\bibliographystyle{alpha}
\bibliography{fot}

\end{document}